\hfuzz 30pt
\magnification \magstep 1
\hsize 6 true in
\vsize 8.5 true in
\input amssym.def
\input amssym.tex
%\nopagenumbers
\openup 1 \jot
\centerline {\bf Tunnelling  with a Negative Cosmological Constant}
\vskip 1.5 cm
\centerline {G.W. GIBBONS}
\centerline {D.A.M.T.P.}
\centerline {University of Cambridge}
\centerline {Silver Street}
\centerline {Cambridge CB3 9EW}
\centerline {U.K.}
\vskip 1.5cm
\centerline  {\bf ABSTRACT}
\tenrm
{\narrower \narrower. \smallskip The point of this paper is see what light
new results in hyperbolic geometry may throw on gravitational entropy and whether 
gravitational entropy is relevant for the quantum origin of the univeres.
We introduce some new gravitational instantons 
which mediate the birth from nothing of  closed universes containing 
wormholes and suggest 
that they may contribute to the density matrix of the universe.
We also discuss the connection between their gravitational action and
the 
topological and volumetric entropies introduced in hyperbolic geometry.
These coincide for hyperbolic 4-manifolds, and increase with 
increasing topological complexity of the four 
manifold. We raise the questions of whether the action also  
increase with the topological 
complexity of the initial 3-geometry, measured either by its three volume
or its Matveev complexity. We point out, in distinction to the 
non-supergravity case,
 that universes with domains
of negative cosmological constant separated by 
supergravity domain walls cannot be 
born from 
nothing. Finally we point out that our wormholes provide examples of
the type of Perpetual Motion machines envisaged by Frolov and Novikov.
}

\beginsection Introduction 

There has been great  interest recently in  calculations
of the semi-classical tunneling rates for the production of pairs of 
black holes in quantum gravity [1] . A notable feature of  these results is
light they have thrown upon the Bekenstein-Hawking entropy 
$$
S_{\rm Bekenstein-Hawking} = { 1\over 4} A
$$
of non-extreme event horizons. 
 
It has been found that the probability of creating 
near-extreme black holes compared with extreme, solitonic, holes
for which $S_{\rm Bekenstein-Hawking}=0$ is increased by a factor
$$
\exp ({ 1\over 4 }A)=\exp (S_{\rm Bekenstein-Hawking} ).
$$

This lends further support to the interpretation of $S_{\rm Bekenstein-Hawking} $
as  a purely gravitational contribution to the total thermodynamic
entropy of any system containing black holes. 

Similar results hold for the entropy $S_{\rm Cosmological}$ of 
Cosmological Horizons. The Euclidean action of $S^4$ and $S^2\times S^2$
is given by
$$
I_{\rm euc}=-S_{\rm Cosmological} = { 1 \over 4} A_{\rm C}
$$

An interesting question is whether there are other circumstances in which
one may associate entropy with other types of gravitational fields.
In particular it is tempting to apply the idea of 
entropy to the intial conditions of the universe. In this paper
I shall seek to do so using hyperbolic geometry
this by considering semi-classical tunneling
models
in which the universea is "born from nothing".  This is
a situation which resembles rather closely the case of pair
production and so one may adopt similar methods. The spatial
sections $\Sigma$ of the Lorentzian spacetimes $M_L$  that
I shall consider 
are hyperbolic 3-manifolds. These are not simply connected and
the fundamental group $\pi_1(\Sigma) $ contains elements of infinite order
(the first Betti-number $b_1(\Sigma) >0$ ) . Thus in a sense 
one may speak of the 
creation of  wormholes. However it should be stressed
that we are not considering connected sums of copies
of $S^1 \times S^2$  so probably one shouldn't think of these
as Wheeler type wormholes, for which there is evidence that 
they have an associated entropy.  We shall  be concerned, in part, with
the question of whether this other type of  
wormhole has an  associated entropy.

\beginsection Real Tunnelling Geometries

Current  models of the quantum origin of the universe begin 
with  a 
"real tunneling geometry" [2], that is a solution of the 
classical Einstein equations which consists of a Riemannian manifold
${ M}_R$ and Lorentzian manifold ${ M}_L$ joined across a 
totally geodesic spacelike surface $\Sigma$. Each connected component 
$\Sigma^i$ of the surface $\Sigma$ serves both as a Cauchy surface for a  
totally disjoint Lorentzian universe ${ M}_L^i $ and a connected component 
$\partial { M}_R^i$ of the boundary $\partial { M}_R$ 
of the Riemannian manifold ${ M}_R$. 
In cosmology $\Sigma$ is taken to be closed (that is compact without 
boundary) and in accordance with the No Boundary Proposal [3] one usually 
takes the Riemannian manifold ${ M}_R$
to be connected, orientable and compact with sole boundary $\Sigma$. 

 One sometimes
sees semi-classical calculations of topology changing amplitudes
with a  boundary which is not totally geodesic.
For example one might remove a number of solid 4-balls out of flat 4-torus
$T^4$ or a round four sphere $S^4$. In the former case the euclidean action
$I_{\rm euc} $ comes entirely from the boundary term and is given by
$$
I_{\rm euc}= \pm { 3 \pi \over 4} R^2
$$   
where the minus sign gives the action of the 4-ball $B^4$ 
of radius $R$ 
and the plus sign gives  the action of $T^4 -B^4$ the 4-torus with 
a 4-ball removed. Removing more 4-balls will evidently increase the action.
Consistent with this Carlip [4]  has shown that the wave function, 
considered as 
a function of both the metric and the second fundamental form
 is stationary if 
second fundamental form vanishes.
One might anticipate that taking
a non-vanishing second fundamental form  would lead to a higher classical
eulclidean action and hence a lower probability. This is certainly 
consistent with the examples above. Thus in what follows I shall assume 
that the 
dominant contributions do indeed come from manifolds with totally 
geodesic boundaries. I will comment on the physical significance
of having more than one boundary component in a later section.

Given this set up one one may pass to the double $2{ M}_R= { M}_R ^+ \cup 
{ M}_R ^-$ by joining two copies of ${M}_R$ across 
$\Sigma$. This is a closed orientable Riemannian manifold admitting a 
reflection map, that is an orientation reversing involution, $\theta $ say 
which fixes the 
totally geodesic submanifold $\Sigma$ and permutes the two portions 
${ M}_R^\pm$. The involution $\theta $ plays a crucial role in the quantum 
theory
because it allows one to formulate the requirement of "Reflection Positivity" (see [5] for details). 

Conversely if one is interested in finding a compact Riemannian manifold
$M_R$ with totally geodesic boundary 
one may start with given a closed orientable Riemannian manifold
${ M}$ and  ask whether it admits
 an orientation reversing
involution $\theta$ which fixes 
an embedded  hypersuface $\Sigma _\theta $ (i.e. one without self-intersections). 
If so
then the  hypersurface is  necessarilly two sided and  totally geodesic. 
One may then cut the manifold along $\Sigma_\theta $. There are now two possibilities. If 
$\Sigma _\theta $ {\sl separates} $M$ then it will, as it were, fall into two 
disjoint 
isometric pieces $M_R^\pm$. This happens in all the cases considered in [2]
including the archetypal case when $M$ is the round metric on $S^4$ and
$\theta$ is reflection in an equator. 

If, on the other hand, 
$\Sigma _\theta $ does not separate then cutting $M$ along $\Sigma _\theta$ will result in a single connected manifold $M_R$ whose boundary $\partial M_R$ is totally geodesic and consists of two disjoint
copies  $\Sigma _\theta ^\pm $ of $\Sigma _\theta $. In this case the involution $\theta $ will act on $M_R$ permuting the two portions of the boundary  $\Sigma _\theta^\pm $.
One may of course now join two copies of $M_R$ together across $\partial M=
\Sigma _\theta^+ \sqcup \Sigma _\theta^-$ to obtain the closed double  
$ M^ \prime =2 M_R$
on which some other involution $\theta^\prime$ acts. Clearly $M^\prime$ is a double cover of the original closed manifold $M$.
In the case that $\Sigma_\theta $ fails to separate the boundary $\partial M_R$ is never connected even though the fixed point set $\Sigma_\theta$ may be connected.     
The construction we have just given with its two 
possible variants really only requires a two sided totally geodesic hypersurface $\Sigma$. 
It need not necessarily be the fixed point set of an 
involution. Given $\Sigma$ we may always cut the manifold $M$ along it.
However finding a totally geodesic hypersurface may be quite 
hard. The easiest way do do so in practice is to look for the fixed point set
of an involution.

Note that there is a connection between the failure of $\Sigma$ to separate
and the topology of $M$ [6]. If a two-sided hypersurface $\Sigma$, 
totally geodesic or not, 
fails  to separate then it cannot bound and thus it represents an non-trivial
homology class in $H^{n-1}(M; {\Bbb R})$, where $n$ is the dimension of $M$. 
 It follows from Hodge duality
that the first Betti number $b_1(M)$ of $M$ cannot vanish. It is well known that
if the Ricci-tensor of   
$M$ is non-negative then the first Betti number must vanish. 
This fact was used in [2] to argue that in this case any boundary must be 
connected: the birth of disjoint Lorentzian unverses is not allowed.
Put another way: if the Ricci tensor is non-negative then the assumptiom made in the No Boundary Proposal that there is only one boundary is redundant :
it follows from the compactness of $M_R$.
It was also pointed out that if the Ricci tensor is not non-negative then 
it is easy to find examples with two boundary components.
In the explicit examples considered in [2] the failure of $\Sigma$ 
to separate was not encountered because they had positive cosmological
constant. I shall comment later on the possible significance of manifolds with more than one boundary component.

Perhaps because a closed Riemannian manifold with negative Ricci
 curvature cannot
admit a Killing vector field there are  few explicitly known examples.
The simplest case to consider the case when the metric is of constant
curvature. This gives rise to a locally isotropic F-R-W " $k=-1$ " universe 
after tunnelling and so is of obvious cosmological interest. 
A closed  4-manifold of constant negative curvature, also referred to as a hyperbolic metric  is of the form
$H^4/\Gamma$ where $\Gamma \subset O(4,1)$ acts properly discontinously and has no parabolic elements. As we shall see in more detail later, it is a theorem [7] that any Einstein metric on the same manifold must also be of constant negative curvature so the 
instantons are unique. This should be contrasted with the more
freqently studied case of the 4-sphere. It is not known whether it admits
Einstein metrics other than the round one. However the proofs of 
the cosmic no hair cosmic theorem  [8]
indicate that the round metric is the only Einstein metric with a 
hypersurface orthogonal circle action.

To date the only explicit attempts to construct tunnelling geometries  known to me have been due Ding, Maeda  and Siino [9].  They glued together 12 eight-cells ( 4-polytopes bounded by 8 congruent hexahedra) to obtain a non-compact
hyperbolic manifold $M_{\rm D-M-S}$ of finite volume with 16 totally geodesic boundary
hyperbolic  components. However the manifold $M_{\rm D-M-S}$ is non-compact and has cusps. The boundary $\partial M_{\rm D-M-S}$ is also non-compact.
They also discuss a similar construction with 16-cells and 24-cells.  

One might wonder whether the cusps are essential. One knows
that Anti-De Sitter spacetime is semi-classically stable [10]. 
This may be proved using the fact that this spacetime is supersymmetric, admitting Killing spinors [10]. One might think that this would rule out the 
spontaneous creation of
closed universes, without cusps. However one the identifications needed
to produce a closed universe  are presumably incompatible with
the existence of Killing spinors and so perhaps. 
This is equivalent to asking whether the No Boundary proposal is 
compatible with a negative cosmological constant. 
supersymmetry is not relevant in this situation.  
One might also ask whether the 
creation of a {\sl single} universe is possible. This is equivalent to
asking whether the No Boundary proposal is compatible with a negative cosmological constant.

 \beginsection New Examples 

To answer these questions one needs to examine more examples.
One without  without cusps is provided by taking  $M= M_{\rm Davis}$ where
 $M_{\rm Davis}$ is a compact orientable hypebolic manifold 
which is obtained by suitably identifying the 120 dodecahedric faces of 
a certain hyperbolic Coxeter 4-polytope $X^4 \subset H^4$ [11] . One has 
$$  
M_{\rm Davis}= H^4/K
$$
where $ K \subset G_4$ and $G_4$ is the Coxeter group generated by reflections in the faces of $X^4$. 
 Translating $X^4$ under the action of $K$ gives a tesselation of $H^4$ by identical regular polytopes -- a so-called non-euclidean honeycomb. The group $K$ is a subgroup of $G_4$ which acts freely on $H^4$ and which is generated by reflections which identify opposite faces of $X^4$.

In what follows we reproduce Davis's description, adhering  to his notation. Basically " it's all done by mirrors ". We may think if we wish of $H^4$  as the mass shell or future spacelike hyperboloid $Q$ in five-dimensional
Minkowski spacetime ${\Bbb R}^{4,1}$. Timelike hyperplanes intersect $H^4$ in hyperbolic planes and each such plane is totally geodesic. A reflection is a reflection in  a timelike hyperplane
and is contained in  $O_{\uparrow}(4,1)$ the group of time-orientation preserving Lorentz transformations in ${\Bbb R}^{4,1}$. The faces $D$ of the poytope $X^4$
are of course planar. Let $r_D$ be reflection in the face $D$.
These reflections generate $G_4$. The polytope $X^4$ is centro-symmetric so join the centre  of the face $D$ to the centre $x_4$ of the poytope by a geodesic and let $s_D$ be reflection across the orthogonal hyperplane through $x_4$.  Clearly $s_D$ takes $D$ to the opposite face $-D$. Let $t_D=r_D s_D$. 
Now $s_D \in G_3 \subset G_4$ where
$G_3$ is the stablizer of the origin $x_4$. Thus $t_D$ also takes $D$ to its opposite face and belongs to $G_4$. The group it generates is $K$. Acting on $X^4$, $t_D$ takes it to the polytope adjacent to $D$ in the tesselation.
   Davis shows that $G_4$ is the semi-direct product of $G_3$ and $K$
and that $K$ acts freely on $H^4$. All elements of $K$, being the products of an even number of reflections, preserve orientation and so the quotient $M_{\rm Davis}= H^4/K$ is orientable. Since $K$ is a normal subgroup
of $G_4$ the quotient $G_3= G_4/K$ acts on $M_{\rm Davis}$. Thus
$s_D$ is an orientation reversing isometry of $M_{\rm Davis}$ 
which fixes a connected totally geodesic
2-sided hypersurface called $M^3$ by Davis and $\Sigma$ here. Moreover, as he points out, the the complement of $\Sigma$ in $M_{\rm Davis}$ is obviously
connected, or
in other words $\Sigma$ does not separate. In fact $1 \le b_1(M_{\rm Davis}) \le 60$. The Euler characteristic is $26$. The second Betti number $b_2(M_{\rm Davis})= 2 ( 12 + b_1(M_{\rm Davis}))$ is therefore even, which is 
consistent with the fact that the Hirzebruch signature
 $\tau= b_2^+-b_2^- $ must vanish because the involution $s_D$ will pull back self-dual harmonic forms to anti-self-dual harmonic forms.

The upshot of all of this is that  cutting $M_{\rm Davis}$ along $\Sigma$ will give a manifold with
a totally geodesic boundary with two connected components.  

Ratcliffe and Tschantz have given examples of non-compact hyperbolic 4-manifolds of finite volume [12]. We shall refer to the simplest example as $ M _{\rm Ratcliffe-Tschantz}$. It is obtained by identifying the faces of a 24-cell.
The vertices of this polytope lie on the absolute at infinity. 
They correspond to the following 24 lightlike vectors in ${\Bbb R}^{4,1}$ : 
$(\pm 1,0,0,0,+1)$, 
$(0, \pm 1,0,0,+1)$, $(0,0, \pm 1,0,+1)$,$(0,0,0, \pm 1,+1)$, $(\pm { 1\over 2},\pm { 1\over 2},\pm { 1\over 2},\pm { 1\over 2}, +1) $.

The polytope is invariant under $O_{\uparrow} (3,1; {\Bbb Z})$ the group of 
integer valued Lorentz transformations preserving the time orientation.
The congruence 2 subgroup  $\Gamma \subset O_{\uparrow} (3,1; {\Bbb Z}) $ consisting of integer valued  Lorentz transformations  congruent modulo 2 to the identity is torsion free (i.e. has no subgroups of finite order) and thus
acts freely on $H^4$. One has   
$$
 M _{\rm Ratcliffe-Tschantz} = H^4 /\Gamma
$$

Evidently the 24-cell
is invariant under the reflection 
$ \theta : {\Bbb R} ^{4,1} \rightarrow {\Bbb R} ^{4,1}$ 
sending $ (X^1, X^2, X^3, X^4, X^0,)$ to $(-X^1, X^2, X^3, X^4, X^0,)$.
The reflection $\theta$ normalizes $\Gamma$ in $ O_{\uparrow} (3,1; {\Bbb Z})$ and therefore descends to the quotient $  H^4 /\Gamma$. Clearly $\theta$ fixes a 
connected totally geodesic submanifold $\Sigma_\theta$  in $ M _{\rm Ratcliffe-Tschantz}$ which
does not separate. Cutting $ M _{\rm Ratcliffe-Tschantz}$  along  $\Sigma_\theta$ therefore yields a manifold $M_R$ with two boundary components.

\beginsection Topological and Volumetric Entropies and The Einstein Action

Formally one may attempt to evaluate the functional integral over all Riemannian metrics on closed 4-manifolds in Euclidean quantum gravity is  
dividing the metrics into conformal equivalence classes. In each 
equivalence class find a representative with with constant Ricci scalar. The integral is then split into an integral over the 
conformal deformations of that representative and an integral over conformal equivalence classes. The integral over conformal deformations must be treated differently because the Euclidean action:
$$
I _{\rm eu} = -{ 1 \over 16 \pi} \int _M \sqrt g d^4 x \Bigl ( R-2 \Lambda \Bigr )
$$

in those directions is not bounded below [13]. If the Ricci scalar $R$ is scaled to take the constant value $ 4\Lambda$ the euclidean action  is 
proportional to its volume $V= {\rm vol} (M,g)$:
$$
I _{\rm eu} = - { 1 \over 8 \pi} \Lambda V.
$$

If the cosmological constant is positive then it is not excluded that
$2M_L$ is locally static, with a Killing horizon of total area $A_{\rm C}$.
If all connected components of the  Killing horizons have 
the same surface gravity $\kappa$ one may anaytically continue to obtain a closed Einstein manifold admitting a reversible circle action.
It then  it follows form the Einstein equations that
$$
{ 1 \over 8 \pi} \Lambda V= { 1\over 4} A_{\rm C}.
$$
The only two known cases known $S^4$ with $\Sigma \equiv S^3$ and one 
component with area
$$
A_{\rm C}= {12\pi \over \Lambda}
$$
and $S^2 \times S^2$ with $\Sigma \equiv S^1 \times S^2$ and two equal
components with total area
$$
A_{\rm C} = { 8\pi \over \Lambda}.
$$

 If the cosmological constant is negative then no static Lorentzian 
Einstein metric
can be anaytically continued to give a closed Riemannian manifold
 and so Hamiltonian methods cannot be used in a straightforward
way to relate
the the action to the gravitational entropy\footnote{*}{ We shall describe in more
detail the Lorentzian sections of hyperbolic manifolds in more detail in 
a later  section}. However for hyperbolic 
manifolds
( i.e. those admitting a metric, call it $g_0$,  of constant 
negative curvature)
the 4-volume and hence the action is known to be related to the 
topological entropy 
$h_{\rm top}(g)$ of the geodesic flow on the unit tangent bundle. 
This suggests that there might be some connection between
topological entropy and gravitational entropy.   

Recall [7] that the definition of $h_{\rm top}(g)$ is
$$
h_{\rm top}(g) = {\rm Lim } _{L \rightarrow + \infty} { 1 \over L} \log ( \# \{ \gamma : l_g(\gamma) \le L\} )
$$
where $l_g(\gamma)$ is the length of the periodic geodesic $\gamma$ with respect to the metric $g$.  
One may also define a volumetric entropy $h_{\rm vol}(g)$
by
$$
h_{\rm vol} (g) = {\rm Lim } _  {L \rightarrow + \infty} { 1 \over L} \log ( {\rm vol}(B(x,L))
$$
where $   {\rm vol}(B(x,L)) $ is the volume, with respect to the metric $g$
of a ball of radius $L$ centred at the point $x$ in the universal covering space $\tilde M$ of the manifold $M$. 

In other parts of physics
or mathematics  one thinks of 
entropy as a convex function on a space $\cal S$ of mixed  "states". For this to make sense the space of states $\cal S$ must be a convex set. 
This is certainly true for a classical probablity distribution
on a finite set when $\cal S$  a simplex,and for its quantum mechanical 
generalization, the set of density matrices for a finite dimensional Hilbert space. The set of Riemannian metrics ${\rm Riem }(M)$
 a compact manifold, unlike the set of Lorentzian metrics  is certianly a convex set. \footnote {*} { Of course this will not remain true once we have taken the quotient by the action of the diffeomorphism group ${\rm Diff}(M)$.} 
It is therfore very striking that Robert [14] has shown that
that the volumetric entropy is a convex function of the metric $g$. 
 
It is known  that topological entropy is always smaller than volumetric entropy:
$$h_{\rm vol} (g) \le h_{\rm top}(g)$$
and that if $g$ has negative curvature then $h_{\rm vol} (g)=h_{\rm top}(g)$. In the case of a hyperbolic metric $g_0$ one has
$$
h_{\rm vol}(g_0) =h_{\rm top}  ( g_0)= \sqrt { -3 \Lambda}.
$$
Note that unlike gravitational entropy which has the dimensions
of area these "entropies"
have the dimensions of inverse length. Moreover Anti-De-Sitter spacetime has
no gravitational entropy. This should be contrasted with De-Sitter
space which in the guise of the 4-sphere has gravitational entropy but has no topological or volumetric entropy. Thus it seems, superficially at least, that these two concepts of entropy are physically unrelated.  
On the other hand, mathematically, gravitational entropy is related to the gravitational 
action. In fact  topological and volumetric entropy are also related to the 
volume of the manifold and thus all three entropies are therefore related  to the action. The connection between event horizons and hyperbolic geometry
will be expanded upon in a later section. 

To see this connection in more detail  recall from [15]
that if the Ricci curvature
has a positive  lower bound:
$$
R_{\alpha \beta } v^\alpha v^\beta \ge |\Lambda | g_{\alpha \beta} v^ \alpha v^ \beta 
$$
then Bishop's theorem tells us that
$$
{\rm vol} (M,g) \le {24 \pi ^2  \over \Lambda ^2} 
$$
with equality if and only if $g$ is the round metric on $S^4$.
Thus the round 4-sphere  has the largest volume and hence the lowest action among all  
metrics with positive cosmological constant.  

On the other hand if one has has a negative lower bound for the Ricci curvature:
$$
R_{\alpha \beta } v^\alpha v^\beta \ge- |\Lambda | g_{\alpha \beta} v^ \alpha v^ \beta 
$$
then one may apply Bishop's comparison theorem to to a ball in the universal
cover $\tilde M$
to obtain  an upper bound for the volumetric entropy and hence an upper bound for the topological entropy:
$$
h_{\rm top} (g) \le h_{\rm vol} \le \sqrt { -3 \Lambda}.
$$ 

Moreover if   a closed 4- manifold $M$ admits a metric $g_0$ of constant
negative curvature, and if $g$ is any other metric on $M$. One has [7]
$$
{\rm vol} (M,g) h_{\rm vol}(g) ^4 \ge {\rm vol} (M,g_0) h_{\rm vol}(g_0) ^4. 
$$

If the Ricci-curvature has a negative lower bound it follows that
the volume is always greater than that of the hyperbolic metric on $M$: 
$$
{\rm vol} (M,g) \ge {\rm vol} (M,g_0).
$$
Now if the metric $g$ is an Einstein metric  the Gauss-Bonnet theorem 
tells us that the volume $V={\rm vol}( M,g_0)$ is given in terms of the
Euler characteristic $e(M)$  by 
$$
V=  { 12  \over \Lambda ^2} \pi ^2 \Bigl ( e(M) -  { 1 \over 32 \pi ^2 }\int _M C _{\alpha \beta \gamma \delta} C^{\alpha \beta \gamma \delta} \sqrt g d^4x       \Bigr  )
$$
where $C_{\alpha \beta \gamma \delta }$ is the Weyl tensor. 

Thus for a hyperbolic metric the volume
$$
V= { 12  \over \Lambda ^2} \pi ^2 e(M).
$$ 

Moreover it follows that for any other Einstein metric on $M$ with the same 
cosmological constant that it's volume is bounded above and below by the same value. This can only happen if the Weyl tensor vanishes and hence it must have constant curvature. This is the uniqueness theorem of Besson {\it et al.} referred to earlier.

The conclusion is that hyperbolic metrics have the largest volumetric  and topological entropy among all metrics on the same manifold and among metrics on the same manifold with Ricci-curvatures having a 
negative lower bound they have
the least volume. Finally and most importantly physically: among all metrics with constant Ricci scalar $R=4 \Lambda$ having a negative lower bound for the Ricci-curvature the hyperbolic metric has the least action.
Note that hyperbolic metrics are {\sl locally} homogenous but not
globally so. In this repect at least, it seems reasonable to think of them as having
high entropy.

The relationship between euclidean action and volume  given above follows 
directly and straighforwardly if one assumes that the cosmological constant is a fixed constant
which remains constant under Wick rotation. If, however,
the cosmological constant is related by duality
to the vacuum expectation value of a closed four-form $ F_{\alpha \beta \gamma \delta}$ the derivation is more subtle . Thus if we add to the Lorentzian
 action a term

$$
-{ 1 \over 48} \int d^4x F_{\alpha \beta \gamma \delta} F^{\alpha \beta \gamma \delta} \sqrt {-g}.
$$

The sign in the action is chosen so that $F$ contributes positively to the energy. It is the sign which would arise naturally arose from a dynamical four-form in higher dimensions. Thus if $c$ is a constant, $ \eta_{\alpha \beta \gamma \delta}$
 the covariantly constant volume form on  necessarily orientable manifold and 
$$
F_{\alpha \beta \gamma \delta}=c\eta _{\alpha \beta \gamma \delta},
$$
then the Lorentzian field equations obtained by varying with respect to the metric $g_{\alpha \beta}$ contain a positive cosmological term with
$$
\Lambda = { 1\over 4} c^2. 
$$
That is they have as a solution De-Sitter spacetime. On the other hand if a term
$$
+{ 1 \over 48} \int d^4x\sqrt {g}  F_{\alpha \beta \gamma \delta} F^{\alpha \beta \gamma \delta} 
$$
is added to the Riemannian action the field equations contain a negative cosmological term, i.e. would have $H^4$ as a solution with $F_{\alpha \beta \gamma \delta } $ real. It may ultimately be significant that if the manifold one is working is non-orientable then one cannot induce a cosmological term in this way.

The Riemannian  solution may be obtained from the Lorentzian one by making both the time and the integration constant $c$
pure imaginary. On the other hand just analytically continuing in the 
time would give a purely imaginary four-form and hence $S^4$ as a solution.
The question then arises: what is the correct instanton solution
and what is Euclidean action
of that solution? A similar question arises when considering the action of electrically charged black holes. In that case experience with black hole thermodynamics supports the idea that the action should be evaluated for a purely
imaginary electric field on the Riemannian section. 
This procedure may be justified {\it a priori} in that case because one wants to evaluate a partition function at fixed real chemical potential. 
An {\it a posteori} justification is that the results so obtained are consistent with electric-magnetic duality. 

In the cosmological context Hawking in an attempt to explain the smallness of the cosmological constant  took the $S^4$ solution and evaluated its action [16] .
Thus amounts to following the electromagnetic case and allowing a purely imaginary four-form. It would preclude using hyperbolic metrics.
This procedure was later criticised by Duff [17]. 
If one follows the Duff procedure one finds that that the Euclidean volume
for a hyperbolic manifold with a real four-form would be proportional to
the volume with a negative proportionality constant. Since there exist
hyperbolic manifolds with arbitrarily large 4-volume this procedure would appear to lead to the unsatisfactory result that these are not suppressed in the path integral. Of course the conclusions above depend upon choosing
the negative sign in the Lorentzian action of the four-form. There appear
to be good reasons for this in the context of superstring and hence supergravity theory but if the signs are reversed the conclusions above would 
be reversed.

In the saddle point approximation one usually only 
considers the classical action of the saddle point and compares the actions of different saddle points. In the case of hyperbolic manifolds  the discussion above shows that this reduces to
comparing their Euler characteristics.

The closed hyperbolic manifold with the  the lowest known Euler 
characteristic is the Davis manifold [11]  for which $e(M_{\rm Davis})=26$.
Thus because the totally hyperbolic hypersurface does not separate
the action of the Davis instanton is therefore 
$$
I_{\rm euc}(M_{\rm Davis})  = {39  \pi \over  |\Lambda|}.
$$

 For hyperbolic manfolds with cusps the volume is given by the same formulae.
It is known that all positive integer values occur. The lowest volume and hence
presumably lowest action manifold with cusps is the example of Ratcliffe and Tschantz [12] obtained by identifying the faces of a 24-cell. This has 
 $e(M _{\rm Ratcliffe-Tschantz})=1$. 
$$
 M _{\rm Ratcliffe-Tschantz} = H^4 /\Gamma
$$

Evidently the 24-cell
is invariant under reflection $ \theta : {\Bbb R}^{4,1} \rightarrow {\Bbb R}^{4,1}$ 
sending $ (X^1, X^2, X^3, X^4, X^0,)$ to $(-X^1, X^2, X^3, X^4, X^0,)$.
The reflection $\theta$ normalizes $\Gamma$ in $ O_{\uparrow} (3,1; {\Bbb Z})$ and therefore descends to the quotient $  H^4 /\Gamma$. Clearly $\theta$ fixes a 
connected totally geodesic submanifold $\Sigma_\theta$  in $ M _{\rm Ratcliffe-Tschantz}$ which
does not separate. Cutting $ M _{\rm Ratcliffe-Tschantz}$  along  $\Sigma_\theta$ therefore yields a manifold $M_R$ with two boundary components.

The volumes of the three solutions calculated numerically
 by Ding, Saeda and Siino [9] 
corresponding to 12 8-cells, 4 16 cells and 6 24 cells give   effective Euler characteristics of 6.2017219, 2.6666776
and 26.993285 respectively. The fact they they are not integral is  puzzling.
One would expect the last number to equal 6.
  
The example constructed from Ratcliffe-Tschantz manifold above 
definitely  has lower in action (assuming that cusps do not contribute), since it is built up from just one 24-cell.

\beginsection Product Examples

Real tunneling solutions of the  Einstein equations 4-manifolds with negative cosmological constant $\Lambda$  may also be be obtained by taking  the metric product of a closed 2-dimensional manifold of genus $g$ with constant 
curvature $ - {1 \over |\Lambda|}$ with a compact 2-dimensional manifold
with constant 
curvature $  -{1 \over |\Lambda |}$ with a geodesic boundary. 
These metrics after tunneling give homogeneous but  anisotropic
cosmological models of the form of  products of two-dimensional anti-de-Sitter spacetime with a closed 2-dimensional manifold of genus $g$ with constant 
negative curvature. In general product metrics of this type are relevant to
the possibility of spontaneous compactification. In the present case
one has in mind compactification form $4$ to $2$ spacetime dimensions.

Using the Gauss-Bonnet theorem one finds [18] that the  euclidean action  is
given  in terms of the Euler characteristic 
and the cosmological constant by   
$$
I_{\rm euc} = {  \pi \over   2|\Lambda|} e(M) ,
$$
where the expression for the Euler number is the product of the Euler numbers of the factors. Thus 
$$
I_{\rm euc} = { 2 \pi \over   |\Lambda|} (g-1) (g_{\rm eff}-1), 
$$
and where the area $A$ of the 2-manifold with boundary is given by
$$
A=  {2 \pi  (g_{\rm eff}-1) \over |\Lambda |}.
$$

The action of a product is smaller by a factor 3 than for a hyperbolic manifold with the same cosmological constant and a Euler characteristic. This is curious because,  naively at least it indicates that anisotropic universes should be formed with higher probablity than anisotropic universes.  

The lowest action case is when both factor manifolds  have the lowest possible
genus. Thus set  $g=2$ and think of  a pretzel as a suitably sized 
regular octagon in the hyperbolic plane with opposite edges identified in the opposite sense. One may cut the pretzel along a geodesic joining 
the mid-points of a pair of opposite edges. This geodesic will not separate
and so $g_{\rm eff}=2$, and hence

$$
I_{\rm euc} = { 2 \pi \over   |\Lambda|}  .
$$

If one takes a  separating geodesic the action would be at least halved.
In either case it is much less than the Davis example.  This is because
the latter has such a high Euler characteristic.

\beginsection Disconnected Boundaries and Density Matrices 

In this section I wish to discuss the physical significance of 
more than one boundary component. If the components are not isometric then
the obvious interpretation is that they give tunneling amplitudes beteween
different three manifolds. From the Riemannian point of view it is not
obvious which components are to be taken to lie
in the future and which to lie the past. 

If we can identify some of the boundary components however a different 
interpretation is possible, as pointed out in a slightly different 
context by 
Hawking and Page [19]. Suppose for simplicity we have two isometric
boundary components. We may glue the four manifold together across 
them to obtain a closed manifold containing a totally geodesic hypersurface
$\Sigma$ 
which does not separate. This is the case for the Davis manifold 
$M_{\rm Davis}$ for example. 

Hawking and Page suggested that one now focus on the "probablity
${\rm Prob}(\Sigma)$ for 
the occurrence of the Riemannian 3-manifold $\Sigma$ ".
This may be   expressed as a functional integral over all 
metrics on all closed 4-manifolds
containing $\Sigma$:

$${\rm Prob}(\Sigma) = \sum_ { M} \sum _g   \exp- I_{\rm euc}(g).
$$
  
The sum decomposes into a sum of terms of the following three
kinds

\item {i} Manifolds $M$ which are separated into two diffeomorphic halves
$M_{\pm}$.

\item {ii} Manifolds 
for which $\Sigma$ separates M into two halves $M_1$ and $M_2$
which are not
diffeomorphic 

\item {iii} Manifolds for which $\Sigma$ does not separate.
  
Terms of the first and second kind have an obvious interpretion in terms
of the Hartle-Hawking type pure state $\Psi_{\rm Hartle-Hawking }(\Sigma)$
$$
\Psi_{\rm Hartle-Hawking }(\Sigma) = \sum_ { M=\partial \Sigma} \sum _ g  \exp- I_{\rm euc}(g).
$$

If one might thinks of the them as the diagonal element of a  
factorized density matrix $\rho _{\rm H-H}$:
$$
\rho _{\rm Hartle-Hawking }= \Psi_{\rm Hartle -Hartle }(\Sigma) \otimes {\overline \Psi_{\rm Hartle-Hawking}(\Sigma) }
$$
then the remaining terms are a measure of the extent to which 
the " density matrix of the universe $\Sigma$ " fails to factorize.

It is clearly tempting to argue that the Davis manifold $M_{\rm Davis}$
represents a semi-classical contribution
 to the non-factorizable part of the 
density matrix of the universe in a theory with a negative 
 cosmological constant. It makes even more interesting the 
question of whether there are hyperbolic 4-manifolds with a 
single connected boundary.

Another viewpoint is that if we only are interested in a connected 3-manifold $\Sigma$ we should include in the functional integral all manfold which bound $\Sigma$ regardless of whether they have other boundary components or not.
Then we have to sum over the 3-metrics on the other boundary componets.
Presumably this gives a mixed state for the universe. 

This is similar to the use of  the formalism of density matrices  applied
to case of a connected boundary in the case 
of spaces with event horizons, as has been done recently
by Barvinsky, Frolov and Zelnikov [20] . Consider the Schwarzshild 
case. The boundary, for which $\Sigma$ has the topology of an Einstein-Rosen bridge 
$S^2 \times {\Bbb R}$,
is given by values of the imaginary Killing time $\tau =0$ and $\tau=4 \pi$. These give the two halves $\Sigma _\pm \equiv S^2 \times {\Bbb R} _\pm $
of the bridge on either side of the throat 
$r=2M$.
 If one is not interested in what happens on one side $\Sigma_-$
of the horizon one should sum over all  metrics on $\Sigma_-$. 

\beginsection Hyperbolic Geometry and Event Horizons

In this section I would like to explore in more detail
the possible relationship between  hyperbolic geometry and gravitational
entropy. We saw above that any relationship between topological
entropy, volumetric entropy and gravitational entropy is is  at best an indirect one.
Neverthless there does appear to be  a common thread:  the 
fact that if the curvature is negative then geodesics diverge exponentially
fast and the the volume of a ball increases exponentially with the radius.
If the manifold $M$ is compact or possibly, as in the case of the fundamental domain of the modular group $H^2/SL(2;{\Bbb Z})$, merely of finite volume  it is
well known that this exponential divergence leads to ergodic behaviour of the geodesics. The topological and volumetric entropies 
$h_{\rm top}(M)$ and $h_{\rm vol}(M)$ were originally introduced to make the relation
more quantitative.

Consider now a static spacetime $2M_L$ with an event horizon. 
The Lorentzian metric takes the form
$$
ds^2 =- V^2 dt^2 + g_{ij} dx^i dx^j
$$
where the positive function $V$ and the 3-metric $g_{ij}$ on the spatial section $\Sigma$ depend 
only on the spatial
variables $x^i$. For simplicity we assume that the event
horizon is non-degenerate and has single connected component. 
The generaliztion of the follwing remarks to more than one component is 
straight forward. Near the 
event horizon the function $V$
tends to zero on some  2-dimensional totally geodesic submanifold $B$ of
the 3-manifold $\Sigma$ :
$$
V \rightarrow  \kappa ^2 s^2 + \dots 
$$
where $s$ is proper distance from $B$ along $\Sigma$ with respect to the 
metric $g_{ij}$, and $\kappa$, the surface gravity is  constant over $B$.
Let us introduce the optical or Fermat metric:
$$
ds^2_o =f_{ij} dx^i dx^j =  V^{-2} g_{ij} dx^idx^j.
$$
Clearly with respect to the optical metric the horizon $B$ is at infinite
distance. If  $B$ has the intrinsic geometry of a 2-sphere
 then the the optical metric approaches the standard hyperbolic metric on 
$H^3$ with radius equal to $1 \over \kappa$.If the original spacetime is    
De-Sitter spacetime then the optical metric is {\sl exactly} that
of hyperbolic three-space. Even if the the intrinsic geometry of the horizon $B$ is not exactly spherically symmetric, locally as one approaches the 
horizon,
the optical geometry approaches, exponentially fast with respect to optical distance, the geometry near infinity in hyperbolic 3-space. 

This universal feature of the optical geometry of event horizons
( which has been noticed before by many people) is very striking and it is 
tempting to try to relate it to the thermodynamic properties of 
event horizons.
Clearly there is no direct  relationship between horizons and ergodicity
because there is no question of making identifications of the optical 
manifold to render it compact or of finite volume.

What one can say however is  that the classical
loss of information about sources which approach the event horizon
which is the subject of the classical No-Hair theorems may be 
seen in this language
as the  loss of information about sources which recede to 
infinity in hyperbolic space. The hyperbolic geometry leads
to an exponential decrease in the multipole moments observed at finite 
points of hyperbolic space. This is a 
simple consequence of the exponential divergence of geodesics.
In other words looking at black holes
is rather like doing astronomy in a static hyperbolic spacetime. 
Since the classical No Hair theorems are rather well understood
in conventional terms it does not seem worthwhile here 
translating them in detail, line for line, into the language of 
hyperbolic geometry but it is clear that this could be done.

At the quantum mechanical level the exponential increase
of the optical volume as one approaches infinity 
is closely related to the fact the taking into account the 
thermal corrections to the classical entropy of a black hole in equilibrium 
at its Hawking temperature $T= { \kappa  \over 2 \pi }$ gives
rise to a an infinite contribution correponding to a gas of massless
particles in equilibrium at the Hawking temperature. Different authors have
attached different significance to this fact. The large number
of states near the horizon is also believed by some to account for
the loss of information during gravitational collapse. This large 
number of states is of course directly related to the infinite optical
volume. 

Thus again it seems that there is probably no  deep connection between
the entropies used in  hyperbolic geometry and the gravitational entropy
of event horizons.
The simple underlying geometrical reason why both concepts are 
useful is the
exponential divergence of geodesics and of volumes. There does not
seem however to be a physical connection and certainly 
it does not seem possible to  identify these entropies physically. 

In the next section I shall suggest that it may be more useful
to think of $h_{\rm top}(M)$ as a measure of {\sl complexity}
rather than entropy.

\beginsection Entropy Action and  Complexity

One often thinks, intuitively at least, that the increase  of
entropy of an isolated macrosopic system is associated with an increase of disorder. 
One has in mind  the fact that if it is isolated an initially complex systems
almosts always evolve into 
a much simpler system, the time reverse  is rather improbable.
It is tempting
therefore to attempt to relate entropy to some measure
of the complexity of a system so that systems with the largest entropy
have the least complexity. One feature of order complexity or order is
spatial inhomogeneity. Thus any definition of 
complexity should presumably have the property  that it is low 
for homogeneous systems. Similarly one expects macrosopic
systems with the largest entropy
to be spatially homogeneous, and this is certainly true 
for ordinary sytems of particles under
the influence of  short range forces. It is not always true in the 
presence of gravity which is a long range field because it tends to 
favour inghomogeneous systems such as stars. For this reason
it is sometimes felt desirable to include a contribution to
the total entropy of a macroscopic system due to gravity which would 
reflect the tendency towards inhomogeneity.

There are many  problems with connecting in any precise sense
entropy and complexity. Firstly one needs a quantitative measure
of complexity.  One such measure is Kolmogorov's
algorithmic complexity. The is related to the shortest 
computer programme required to specify the system. 
More generally one might hope to quantify  the
amount of information or data need to specify the system.
Indeed many people identify entropy with information
although in the case of black hole this often seems  to 
lead to more confusion  than enlightenment, not least because
of a failure to specify what is meant by information.
If one has in mind a probability distribution,
or in quantum mechanics a density matrix
then indeed ordinary thermodynamic or Gibbs entropy $S_G$
$$
S_G= -{\rm Tr} \rho \log \rho / {\rm Tr \rho} 
$$
and the Shannon information $H_S$  gained if 
one discovers  precisely
what state we are in 
are effectively the same thing. For complex systems on the other 
hand one may need to give a great deal of information or data 
to specify them. An ensemble of  systems, each of which is individually 
complex may 
therefore have a large amount 
of information carrying capacity.  Thus two polarization states
of a gravitational wave have more information carrying capacity
than one and the greater the bandwidth of the 
gravitational waves one considers the greater the 
informatiion carrying capacity. Thus the maximum entropy of an ensemble of 
complex systems should be large.

The difficulties in making the idea of 
gravitational entropy in general precise
are well known. 
They include the problem that  thermodynamic entropy is usually 
associated
not with a single system but a class or ensemble of systems.
In the case of a spacetime with an horizon the ensemble 
is often thought of in some sense as all spacetimes which  are 
identical outside the event horizon. It is thus reasonable to 
attemtpt to identify $S_G$ with $S_{\rm Bekenstein-Hawking}$ or $S_C$.

A gravitational wave on the other hand, provided its amplitude, 
phase and polarization state is known should 
presumably have no entropy associated with it. 
Quantum-mechanically one thinks of it as a single coherent state
not a density matrix. By extension one would anticipate that a general
gravitational field without an event horizon should not possess 
thermodynamic entropy. Nevertheless such a field may be very complex.
It seems reasonable therefore, if only in the interest
of conceptual clarity,  to shift ground somewhat
and try to consider how one might define the complexity
of gravitational fields and then afterwards to see whether 
 complexity is related  to other quantities such as the area  of event 
horizons, the Weyl tensor , 
or the euclidean action. In fact Penrose has tried to relate
gravitational entropy to the Weyl tensor for some time and
Dzhunushaliev tries to equate 
Kolmogorov's algorithmic complexity to the classical euclidean action $I$.
It is
known that for black holes of area $A$ and Bekenstein-Hawking entropy $S$
$$
I_{\rm euc} = - S_{\rm Bekenstein-Hawking}= -{ 1 \over 4} A.
$$

Let us turn then to the  question of  how might we define the complexity
of a gravitational field. One approach might be to ask roughly speaking
how many equations
are needed to specify the gravitational field?  A homogeneous spacetime
$M=G/H$ like De-Sitter spacetime is clearly requres rather little information
in this sense. The same is true of the 4-sphere. 
They also contain little information in a slightly different sense.
They do not contain  much information about
the equations they satisfy. This is because they are stationary points of {\sl any} local 
diffeomorphism-invariant action functional
constructed from just the metric and its derivatives. They are examples
of what Bleeker [23] has called {\sl critical metrics}. He showed that if  
${M,g}$ is a closed Riemanian manifold whose metric $g$ is critical
then $M=G/H$ where the isotropy subgroup $H$ of the isometry group $G$
acts irreducibly on the tangent space. In four dimensions the only critical metrics are ( up
to an constant multiple) the round metric on $S^4$
and the Fubini-Study metric on ${\Bbb C} {\Bbb P}^2$. Of them,
$S^4$ can provide a real tunneling geometry. Thus if we assumed
the universe began in a state of least complexity and we took this to mean
that the double $2M_R$ was a critical metric we are led to 
De-Sitter spacetime as the  Lorentzian portion $M_L$.  Now the 4-sphere has
 by Bishop's theorem the least Einstein action among metrics with constant
positive scalar curvature and Ricci-curvature bounded below by 
a non-negative 
multiple of the metric In particular it has least action among all 
Einstein metrics with positive scalar curvataure. 
This at least goes in the same direction as Dzhunushaliev.
The vanishing of the Weyl tensor is also consistent with Penrose's idea 
in this case. However we shall see shortly that if we consider hyperbolic
metrics then while Dzhunushaliev's idea still seems to work
there are problems with that of Penrose. \footnote {*}{Although there may be global problems with the definition
of the action functional, one could apply Bleeker's idea to 
Lorentzian metrics. For Lorentzian 4-metrics 
there is no analogue of ${\Bbb C} {\Bbb P}^2$ so one would be
left with the maximally symmetric spacetimes among homogeneous
examples. However one also acquires an additional case: the Ricci-flat
p-p waves. The structure of the curvature tensor means that they satisfy
any covariant equations constructed from the metric, Riemmann tensor
and its covariant derivatives. Moreover any invariant built from
the Riemmann tensor, including the square of the Weyl tensor, vanishes. 
The cosmological constant must vanish 
of course, though a generalization exists for negative cosmological constant.
It is rather striking that these metrics are also supersymmetric since 
they admit a covariantly constant spinor field. As noted above these
metrics are quite complex and an ensmble may carry entropy. It seems therefore that for lorenztzian metrics the square of the Weyl tensor cannot be taken in a simple unqualified way as a measure of either entropy or complexity.
Neither does it say much about criticality in Bleeker's sense.}

It seems natural to think of closed hyperbolic 4-manifolds as having higher
complexity, however we define it, than $S^4$, the complexity presumably increasing with increasing Euler number.  They are certainly not critical metrics 
in Bleeker's sense since although they are locally homogeneous there is no global isometry group.. The Einstein action is also larger than that of $S^4$
and increases with Euler number.  This agrees in spirit with Dzhunushaliev's 
proposal but presents a problem for Penrose's suggestion  because 
all of these metrics have vanishing Weyl tensor.  In this sense
at least the Weyl curvature hypothesis would seem
to require supplementing in order to render it unambiguous.\footnote {*}{
We saw above that the Einstein action of an Einstein metric is proportional
to its Euler number. It is perhaps worth pointing out here that  
for K\"ahler 4-manifolds with constant Ricci scalar $4\Lambda$ regardless of whether they satisfy the Einstein equations one has 
$$
{\Lambda V \over 8\pi} \ge {9 \pi \tau (M) \over 4 \Lambda},
$$
where $\tau(M)$ is the Hirzebruch signature of $M$.
One has equality for the case of constant holomorphic sectional curvature
and the trivial case of $S^2  \times H^2$. If the cosmological
 constant is positive one is led to ${\Bbb C}{\Bbb P} ^2$ with the 
homogeneous
Fubini-Study metric for which $\tau(M)=1$. 
Thus although they cannot serve as real tunneling geometries,
and although we have as yet no way of associating gravitational 
entropy with them, we see
that K\"ahler metrics are consistent with the general idea that
complexity ( in this case topological complexity measured 
by the Hirzebruch signature) and Einstein action increase together 
and 
that critical metrics are
associated with the smallest possible action.}

\beginsection Entropy and Complexity of Initial Data

Rather than thinking of the 4-metric one might prefer
to think of the entropy or the complexity of the initial data specifying
the spacetime. In the present case this is just the hyperbolic 3-metric 
induced on $\Sigma = \partial M_R$ with vanishing second fundamental form. 
One obvious approach to defining the entropy of intial data is to consider the areas of any black hole  or cosmological  horizons
apparent horizons on the hypersurface $\Sigma$. 
Because $\Sigma$ is totally geodesic these  are minimal 2-surfaces
lying in $\Sigma$. Black hole horizons correspond to "stable" minimal surfaces, i.e. those whose second variation is non-negative.
The known cosmological horizons (for postive $\Lambda$) have just
one negative mode (i.e. the Hessian of the second variation 
is negative on a one-dimensional subspace, or put in another 
way the Morse index is one.) In the case of static solutions with
positive $\Lambda$ the apparent horizoms coincide with event horizons
and Killing horizons and they are totally geodesic submanifolds.
As mentioned above their area is directly related to the action
and to gravitional entropy. For non-static time-symmetric intial
data it follws from the second variation 
that the topology of a connected black hole apparent horizons
 must be that of a 2-sphere and the 
that its  area $A$ is bounded above by
$$
A \le { 4 \pi \over \Lambda}.
$$
Again if $\Lambda $ is positive it seems very likely (following unpublished work with
S-T Yau, G T Horowitz and S W Hawking ) that the area $A$
of an index one apparent  cosmological
horizon is bounded by:
$$
A \le { 12 \pi \over \Lambda}.
$$

In the case of negative $\Lambda$ one does not 
expect cosmological horizons and none of the proofs above go through.
The topology of stable minimal surfaces  does not seem to be restricted to 
that of a 2-sphere and  even if it is,  judging by the 
Schwarzschild-Anti-De-Sitter solution, 
there is no upper bound to its area. It seems therefore that for negative 
cosmological constant the area of any apparent horizons is not necessrily
and interesting quantity to relate to entropy, action
or complexity . Nevertheless not a great deal 
seems to be known  about
it. More information would clearly be desirable.

By contrast Hayward and Twamley have suggested,
in the context of suggestions that our universe should be spatially closed
 with hyperbolic sections, 
that one take the  3-volume ${\rm vol }(\Sigma)$ normalized to a radius of 
unity
as a measure of the complexity
of the inital data and they expressed the feeling that metrics with high 
complexity should be less probable than those with low complexity. I shall
comment on this point later. Before doing so it may be helpfull
to recall why the volume ${\rm vol }(\Sigma)$ might be regarded as a measure of complexity.
It is known from the work of J\o rgensen and Thurston that the set of volumes is a well-ordered closed subset of the real line and that the number of 
closed hyperbolic 
3-manifolds  with the same volume is finite (but may be arbitrarily large).
Moreover Matveev and Fomenko have conjectured [27] that the 
volume ${\rm vol }(\Sigma)$ grows with $d(\Sigma)$, where $d(\Sigma)$ is a 
topological invariant taking  integer values 
which is additive under connected sum:
$$
d(\Sigma_1 \# \Sigma _2)= d(\Sigma _1) + d(\Sigma _2),
$$
and which was introduced by Matveev to study 3-manifolds and which
he calls complexity. I shall call it the Maveev-invariant. 
It vanishes for $S^3$, ${\Bbb R} {\Bbb P}^3$ and the Lens-space $L_{3,1}$.
Matveev's invariant  exceeds $8$ for hyperbolic manifolds
and takes the value $9$ for $Q_1$, the hyperbolic manifold 
with smallest known volume.

Clearly it would be of great interest know how the volume 
${\rm vol }(\partial M)$ of the totally geodesic boundary of a compact
hyperbolic 4-manifold $M$  varies with the 4-volume ${\rm vol }( M)$. 
If they increase together one would have some sort of vindication
of the idea that Euclidean action and complexity are related. 

Of course an alternative viewpoint might be to regard the "entropy"
$h_ {\rm top}$ as giving a measure of the complexity of the Riemannian 
manifold $2M$. Then  Dzhunushaliev's conjecture amounts to relating
$h_ {\rm top}$ to Kolmogorov's algorithmic complexity.

\beginsection Matter Entropy

The calculations so far have been concerned with a vacuum gravitational 
field and hence the entropy or complexity has been purely gravitational. 
In a dynamical situatation it is of 
course only the total entropy which is expected to increase.
In cosmology one anticipates  a sort of competition between gravity and 
matter in which the natural tendency of the matter to homogenize,
erasing structure and complexity, is offset by the tendency of 
gravity to produce inhomogeneities and hence complexity by such
mechanisms as the Jean's instability.

The inclusion of matter has three  effects. Firstly, and most obviously,
assuming that the typical wavelengths of the matter are large compared 
with the radius of curvature, the local contribution to the entropy from the matter must be included
in the total entropy. Secondly, and more subtely, the effect of the matter on the background gravitational field,
and in particular the volume of space, must be taken into account. Thirdly,
if the matter temperature is very low, non-local Casimir-type effects 
due to the geometry of spacetime 
will affect the entropy of the matter. Under this heading 
I would include effects due to horizons. 

A rather simple but illuminating model, which ignores the third effect, 
is obtained by considering a perfect radiation fluid with pressure
equal to one third of the energy density $\rho$ [28]. 
As before the cosmological constant $\Lambda $ is considered fixed.
The Friedman equation
tells us that for a F-L-R-W universe the scale factor $a(t)$ satisfies 
$$
{\dot a}^2= -k +{ \Lambda a^2 \over 3} + { 8 \pi \over 3} {\rho_0 \over a^2}
$$
where
$$
\rho = { \rho \over a^4}.
$$
 
The constant $\rho _0$ is related to the conserved total matter entropy 
$S_{\rm matter}$ :
$$
S_{\rm matter} \propto \rho _0 ^ { 3 \over 4} {\rm vol} (\Sigma)
$$
where the constant of proportionality depends upon the composition of
 the matter. In general there exist initial data with 
arbitrarily large (or small ) entropy. However 
if $k$ and $\Lambda$ are both positive the matter entropy $S_{\rm matter}$
of initial data admiting a moment of time symmetry is bounded
above by the value it takes for ESU, the  Einstein Static Universe [28] .
The least value of $S_{\rm matter}$ under these circumstances is of course zero 
which corresponds  to the empty De-Sitter universe. 

One may consider  inhomogeneous time symmetric intial data.
The Einstein Static Universe turns out to be a local maximum of the 
matter entropy functional $S_{\rm matter}$ [28] .  For fixed volume it as always 
entropically
favourable, by Jensen's inequality for the matter to be homogeneous. 
If the volume is allowed to 
vary this still remains true for radiation. It is not true however
for soft equations of state. If ESU is unstable to the Jeans instability 
then it is not a local maximum for the 
matter entropy functional $S_{\rm matter}$.
 
The situation when both $k$ and $\Lambda$ are negative is different.
There is no upper bound for $S_{\rm matter}$ for time symmetric initial data,
even though the volume ${\rm vol }(\Sigma)$ is bounded.

One might be tempted to regard $e^{S_{\rm matter}}$ as providing
an estimate of  probability of creating matter with these initial data.
In that case if $k$ and $\Lambda$ are both  positive one would 
assign greatest probability to the ESU. If $k$ and $\Lambda$
are both negative there is no data of greatest probablity.
However these estimates would ignore the gravitational contribution.
The gravitional action  is difficult to estimate because there are no
non-singular  Riemannian solutions $\{ M_R, g_R\}$ 
with a single boundary component.
This is because the Riemannian matter entropy current $S_{\rm matter}^\alpha$ 
is divergence free and orthogonal to the boundary $\Sigma = \partial M_R$. 
Therfore it must have a singular point in the interior of any
Riemannian solution. At this point the matter density is infinite.
If one ignores this singularity problem the gravitational contribution
to the action would be proportional to the 4-volume
${\rm vol} (M_R)$ since there is no boundary term. For the case when
both $k$ and $\Lambda$ are both positive  matter and
and gravity  contribute to the probablity with the same sign
and this would favour
overwhelmingly the Einstein Static Universe which has infinite
4-volume. For the case when
both $k$ and $\Lambda$ are both negative matter and
and gravity  contribute to the probablity with the opposite sign.
It is not clear to me which gives the larger effect but in any event 
it seems likely that the probability will be peaked around universes 
resembling Lemaitre's primordial atom.

\beginsection Cusps and Extreme Black Holes

The physical role of cusps is rather obscure. 
However they have been encountered before in connection with extreme
black holes with non-zero horizon area. Near the horizon these metrics are
typically
well approximated by the Robinson-Bertotti solution. This is an exact 
solution
of the Einstein-Maxwell equations the euclidean section of which is
the metric product $H^2 \times S^2$. The radii of curvature are equal in 
the
Einstein-Maxwell case but in more general  examples  this is not true.   

Now imaginary time translations consist of translations along a set of
horocyles
of the $H^2$ factor. Thinking of $H^2$ as the interior of the unit disc
the horocyles are the set of circles passing through a point $p$ at infinity,i.e. a point $p$ on the unit
circle. The constant imaginary time surfaces are geodesics orthogonal to the set of horocycles and
correspond to circles through $p$
which are orthogonal to the unit circle.  If one makes an identification
in imaginary time with a given  period one take just the part of the 
manifold between two of these
geodesics and then identify them.  The result, $H^2/{\Bbb Z}$  is a Beltrami type pseudo-sphere, i.e. a 2-manifold of constant
negative curvature with topology $S^1 \times {\Bbb R}$ and a cusp at infinity.
For visualization purposes it may be isometrically embedded 
in three -dimensional euclidean space as a surface of revolution looking 
like an infinitely
long horn. The surface of revolution is obtained by rotating a
 tractrix curve about its asymptote. The four manifold is just the 
product $H^2/{\Bbb Z} \times S^2$ of the 
Beltrami pseudo-sphere $H^2/{\Bbb Z}$  
with a standard 2-sphere $S^2$.  

It seem that the generic spherically symmetric extreme black hole
with a regular horizon which is identified in 
imaginary time has this structure. It is tempting, therefore,
to regard the cusps encountered in hyperbolic 4-manifolds
as  generalizations of this phenomenon. If this view point is correct
it should be possible to find an analogue among the three-dimensional
black holes. 

\beginsection Supergravity Domain Walls.

If the cosmological "constant" is not really constant but
merely approximately so in a region where a scalar field
is close to one of its vacuum values one may have domain walls.
The simplest case is when the cosmological constant vanishes 
in two symmetric vacua, In the thin wall approximation each domain
corresponds to the interior of a timelike hyperboloid in flat Minkowski
spacetime. The complete spacetime is obtained by gluing two such interiors
back to back across the hyperboloid which represents the history
of the domain wall. The spatial cross sections are diffeomorphic to 
the 3-sphere $S^3$ and  consist
of two flat 3-balls glued back to back.  
The Riemannian section  is obtained by gluing together two flat 4-balls
to obtain a 4-sphere which is almost everywhere flat. 
There is just a ridge of curvature separating the two domains which 
corresponds to  the the history in imaginary time of the domain wall.
The Riemannian section $2M_R$ is invariant under $SO(4)$, and the Lorentzian
section $2 M_L$ under $SO(3,1)$. The nucleation hypersurface $\Sigma$
is compact and is just the two flat 3-balls glued back to back.
The domain walls are repulsive and this makes possible the simultanoues nucleation of black hole pairs [29].

There is clearly a similar construction for negative cosmological constant.
One approach would be glue back to back two hyperbolic 4-balls of finite 
radius. This would give a 4-sphere which has   constant negative
curvature almost everywhere. Again the Riemannian section $2M_R$ is invariant under $SO(4)$, and the Lorentzian
section $2 M_L$ under $SO(3,1)$. The nucleation hypersurface $\Sigma$
is compact and is just the two hyperbolic 3-balls glued back to back.

Another type of domain wall which is {\sl static} and has 
Poincar\'e $E(2,1)$
invariance is also possible and has arisen in supergravity theories [30].
Locally the metric takes the form:
$$
ds^2 = A(z) \Bigl ( -dt^2 + (dx^1)^2 + (dx^2)^2 + (dz)^2 \Bigr ).
$$
If
$$
A(z) = -{ 3  \over \Lambda z^2}
$$
with $\Lambda <0$, then we obtain one half of Anti-De-Sitter spacetime $ADS_4$. 
The horospheric 
coordinates $(t,x,y,z)$ make manifest the Poincar\'e subgroup 
of the full Anti-De-Sitter group:

$$
E(2,1) \subset SO(3,2). 
$$
If we take $0 < z <-\infty$ then Spacelike Infinity is $z=0$ and  $z= + \infty$ is
a null surface through which one may continue the solution
to obtain the complete $AdS_4$ spacetime. The Euclidean section
is obtained by setting $t=i\tau$ with $\tau$ real. One then
obtains the generalized upper half space model of Hyperbolic
4-space
$$
H^4 \equiv {\Bbb R}_+\times {\Bbb R}^3
$$
where ${\Bbb R}_+\times {\Bbb R}^3 = (\tau,x,y,z) $ with $-\infty < \tau < + \infty$, $ -\infty < x < + \infty$, $-\infty < y < + \infty$, and $0 < z < + \infty$.
This construction thus makes manifest the Euclidean subgroup of the full
De-Sitter group:
$$
E(3) \subset SO(4,1).
$$

For the simplest static supergravity domain walls $A(z)$ is an even
function of z which is bounded at $z=0$ and which tends at large $z$
to $ -{3 \over \Lambda  z^2}$. The spacetime looks like  two copies of Anti-De-Sitter spacetime glued together across
spatial infinity $z=0$. Looking  globally one discovers that certain points 
must be omitted. 
The Riemannian section $2M_R$ evidently consist of two copies of the upper 
half space ${\Bbb R}_+\times {\Bbb R}^3 $ glued across the ideal boundary or absolute at $z=0$.
Both geometrically and topologically this similar but
is not the same as the example 
of two hyperbolic 4-balls glued back to back described earlier.
Geometrically this is clear because the isometry groups are different,
$E(3)$ as opposed to $SO(4)$. 
Topologically we now have ${\Bbb R} ^4$ rather than $S^4$ because
the points $\tau^2 + x^2 + y^2 \rightarrow \infty$ are not included.
In other words there is some sort of cusp present.

Physically the most important difference between the two examples is that
while the more or less conventional $SO(4)$-invariant case has
has finite 4-volume and   finite  action the $E(3)$-invariant
supergravity examples have  infinite volume and infinite action. 
This is not just because the nucleation hyper-surfaces $\Sigma$
are of
finite or infinite volume respectively. In the supergravity case
the nucleation surface  $\Sigma$ may be taken to be given by $\tau=0$. 
We could make this have finite volume  by taking by periodically identifying
$x$ and $y$. Even if we did that there would be ( at zero temperature)
be no justification for making the range of $\tau$ finite. 
Since $\partial \over \partial \tau$ is a Killing vector it follows that
the action integrand must be independent of $\tau$ and hence the action
integral over $\tau$ will diverge. 

Thus it sees clear, 
by constrast with the accelerating domain walls,
 that static  domain wall
spacetimes  of this type
cannot spontaneously appear from nothing. In fact we have not used 
any special properties of supergravity in this discussion.
In particular we have not made use of the fact that 
typically examples arising in supergravity satisfy
Bogomol'nyi bounds, 
are partially supersymmetric and admit Killing spinors. Nevertheless
our conclusion is precisely what one would have anticipated 
of such spacetimes.

\beginsection  Lorentzian Sections and Perpetual Motion Machines

One way of describing the  associated Lorentzian manifolds of the Davis example (and others like it) is to use the Gaussian 
coordinate system constructed from the timelike geodesics orthogonal to
$\Sigma$. The result is two connected
copies of 
a F-R-W model with compact spatial sections of constant negative curvature
diffeomeorphic to $\Sigma$, 
 with scale factor $a(T)= \cos( \sqrt { |\Lambda|  \over 3}T)$,
where $T$ is propertime measured along the geodesics:
$$
ds^2 = -dT^2 + a^2(T) g_{ij}(x) dx^i dx^j
$$
where locally $g_{ij}$ is the standard metric on hyperbolic 3-space
$H^3$.. These two
universes begin at
$T=0$ and  collapse to a Big Crunch  at 
$T= {\pi \over 2} \sqrt{ 3 \over |\Lambda |}$. This would be a 
mere coordinate singularity if we were considering
the  covering space, Anti-De-Sitter spacetime, but in our case it 
is a true singularity by virtue of the spatial identifications needed to 
make
$\Sigma$ closed. 

Thus if $|\Lambda |$ were Planck size,  and so the damping effect of 
the action rather small, these topological fluctuations might
not last very long.
However even if $|\Lambda|$ were intially large and hence the initial 
universe rather small,  one could imagine  in  a 
more realistic model that a  small universe formed initially in this 
way might be blown up to macroscopic size by some subsequent inflationary 
process
in which the effective cosmological constant became positive. This might 
mean
 that the non-trivial topology could have observational consequences.
However for this to happen the final size would have to be of the order
of the present Hubble radius and there is no obvious reason why this 
should be the case. Indeed one usually  expects an inflationary period to 
overshoot
so that the characteristic size would be vastly greater than the present 
Hubble radius. For what it is worth, observational searches for the
indications of large scale topology have been rather negative and the  
results of COBE give rather stringent 
limits on  the size of such structures (see [31]  for a recent review
of such models and of  the observational situation). For a recent,
and moore optimistic view of the cosmological significance
of this type of model see [32].

There is another, and in some ways rather more interesting way of 
of describing the Lorentzian sections. 
Since the metric is locally that of Anti-De-Sitter spacetime
which is globally static with Killing time coordinate $t$ say
we have a locally static metric. It is not however globally static
because the identifications made to compactify $M_R$ and
hence $\Sigma$ 
do not commute with the time-translations generated by 
$\partial \over \partial t$.  This sort of situation has been 
discussed by Frolov and Novikov [33] in connection with 
wormholes and time travel. It has a number of interesting consequences
among which is the fact that although energy is locally conserved
it is not globally conserved. We can 
examine this phenomenon in detail in our case.

Locally, in each connected component of $M^i_L$,we may express the metric as 
$$
ds^2 = -V^2(x) dt^2 + g_{ij}(x) dx^i dx^j 
$$
where $g_{ij}$ has the same significance as before and 
where the metric function $V^2$ is a solution of
$$
\nabla ^2V  _g +\Lambda V=0.
$$
The connected initial  hypersurface $\Sigma^i $ for $M^i_L$ may be taken to be at $t=0$.
In other words we have embedded $\Sigma^i$ into a $t=0$ hypersurface
of Anti-DeSitter spacetime. 
Thus  inside a sufficiently small ball
centred  on some point $p \in \Sigma^i$ in spherical coordinates 
one has ( choosing units such that $\Lambda =-3$)
$$
g_{ij} dx^i dx^j = {dr^2 \over 1+r^2} + r^2 \Bigl 
(d \theta ^2 + \sin ^2\theta d \phi ^2 \Bigr)
$$
and
$$
V^2 = 1+ r^2.
$$
Now this coordinate system cannot be extended to arbitrary large radii
because eventually  
it would take us outside $\Sigma^i$. We may think of $\Sigma ^i$
sitting inside $H^3$ as compact subset  with a boundary, 
the points of which are  suitably identified. However there
are points which must be identified at which the metric function
$V$, which gives the length of the 
timelike Killing field $\partial \over \partial t$, 
does not take the same value. Thus clearly the action of time translations
cannot be smoothly extended  over all of $M_L$. 
Normally one thinks of $V$ as
the energy per unit mass of a particle at rest with respect to the 
Killing field $\partial \over \partial t$. Energy conservation demands
that 
as one passes around a closed curve one should get back to the same 
value of the potental energy. In the present situation that cannot happen.
The wormholes that have been created can act as Perpetual Motion Machines
of the Second Kind.
This is precisely the phenomenon described by Frolov and Novikov. Note
however that in our case the wormholes cannot be used as Time Machines.
There are no closed timlike curves in $M_L$. In fact
the coordinate function $T$ will serve as a time function

One may see this more explicitly in the non-compact example of Ratcliffe
and Tschantz. The  24-cell lies in the hyperboloid ${\tilde M}_R$ given by
$X^0 >0$,
$$
(X^0)^2 - (X^1)^2- (X^2)^2- (X^3)^2- (X^4)^2=1.
$$
The totally geodesic surface $\Sigma$ lies in the hyperbolid $\tilde \Sigma$
obtained by setting $X^1=0$. If we introduce an extra timelike 
coordinate $Y^1$, the  3-dimensional hyperboloid $\tilde \Sigma$
may  obtained by setting
$Y^1=0$ in the Anti-DeSitter hyperboloid ${\tilde M}_L$ given by 
$$
(X^0)^2 + (Y^1)^2- (X^2)^2- (X^3)^2- (X^4)^2=1.
$$ 
The tilde  indicates that the relevant
spaces are (non-universal) covering spaces of $2M_R$, $\Sigma$ and $2M_L$.
The two  quadrics ${\tilde M}_R$ and ${\tilde M}_L$ are real slices of the 
complex quadric ${\tilde M}^{\Bbb C}$
$$
(Z^0)^2 -(Z^1)^2-(Z^1)^2-(Z^1)^2-(Z^1)^2 =1.   
$$
which intersect in $\tilde \Sigma$. In fact it is not really necessary to 
consider complexifying $X^2,X^3, X^4$
so lets keep them real. The symmetry group  $G _\Sigma$ 
preserving  $\Sigma$  is
a subgroup of $O_{\uparrow} (3,1;{\Bbb Z}) \subset O_{\uparrow} (4,1;{\Bbb Z})$ the set of integral Lorentz transformations of 
$(X^0, X^1, X^2, X^3, X^4)$ space acting on $H^4$ which leave 
invariant $X^1=0$. 
The same group
$G _\Sigma$ acts by isometries on  the Lorentzian section
${\tilde M}_L$  ${\rm Ad-S}_4$ as a subgroup  
 $O (3,1;{\Bbb Z}) \subset O (3,2;{\Bbb Z})$ of the set of integral 
Lorentz transformations of 
$(X^0, Y^1, X^2, X^3, X^4)$ space which leave invariant $Y^1=0$. Now the 
time translation group $SO(2) \subset SO(3,2)$  of ${\rm Ad-S} _4$ generated by 
$\partial \over \partial t$
corresponds to rotations of the $(X^0,Y^1)$ 2-plane keeping the 
$(X^2,X^3,X^4)$ coordinates fixed. It is  clear that $G_\Sigma$ 
does not commute with the action of  time translations.

\beginsection Attractors and Eschatology

In other parts of physics one frequently encounters 
the claim that there a connection beween basins
of attraction and states of high entropy, since both are related to 
apparently reversible behaviour. 
The usual thermodynamic example is the concept of an equlibrium state.
In the gravitational context this is the intuition behind the various
formulations of dynamical No-Hair and area increase theorems for event 
horizons.
I have nothing new to say about this.
It is of interest however to ask about other gravitational attractors.
In the cosmological case this ammounts to examining the 
solutions of the Einstein equations at very late times.
In that connection it is  of interest to recall that in the context of 
homogeneous Bianchi cosmological vacuum models there is some evidence
that a particular pp-wave, the so-called Lukash solution is an atttractor.
The restriction to spatial homogeneity is probably not necessary
and  on the  grounds that this is just one type of gravitational wave 
one might conjecture that all pp-waves have this property. 
This would be analogous to the behaviour in an asymptotically
flat Minkowski spactime that ultimately all radiation is outgoing.
In the cosmological context it is well known that 
the fact that the universe is not a closed system  
but permits the escape of radiation is why the Universe is not yet
in a state of 
Heat Death. One may ask whether it ever will be. Whatever the answer
it is perhaps fitting to finish this paper with the observation that 
, while in the long run we as
individuals may be dead if the cosmological constant vanishes 
then the universe  will ultimately tend to a supersymmetric state. 
SUSY will live forever!    
 
\beginsection Acknowledgement

It is a pleasure to acknowledge some extremely useful  conservations 
on hyperbolic geometry with Dr A Reid.

\beginsection References

\medskip \item {[1]}
{G.W. Gibbons,
in {\sl Fields and Geometry}, proceedings of
22nd Karpacz Winter School of Theoretical Physics: Fields and
Geometry, Karpacz, Poland, Feb 17 - Mar 1, 1986, ed. A. Jadczyk (World
Scientific, 1986

D. Garfinkle and A. Strominger,
{\sl  Phys. Lett.}  {\bf 256 B } (1991) 146

D. Garfinkle, S. Giddings and A. Strominger {\sl Phys Rev}  {\bf D 49} (1994) 958

S W Hawking, G T Horowitz and S Ross {\sl  Phys Rev} {\bf D 51 } (1995) 4302-4314

C Teitelboim {\sl Phys Rev} {\bf D } (1995) 

 G W Gibbons and R Kallosh {\sl Phys. Rev.} {\bf D51} (1995) 2839-2862 

G W Gibbons  Topology, Entropy and Witten Index of Extreme Dilaton
Black Holes in the Proceedings of the Puri meeeting on Quantum Gravity edited by J Maharana

\medskip \item {[2]} G W Gibbons and J B Hartle {\sl Phys Rev} {\bf D 42} (1990) 
2458-2468  

\medskip \item {[3]} J B Hartle and S W Hawking {\sl Phys Rev} {\bf D28} (1983) 2960-2975 

\medskip \item {[4]} S Carlip {\sl Classical and  Quantum  Gravity } {\bf 10} (1993) 1057-1064 

\medbreak \item {[5]} G W Gibbons: Topology change in classical and 
quantum gravity 
{\sl Recent
Developments in Field Theory} ed. Jihn E Kim ( Min Eum Sa, Seoul ) (1992).

G W Gibbons: Topology change in Lorentzian and Riemannian Gravity,  

{\sl Proceedings of the Sixth Marcel Grossmann Meeting: 

Kyoto 1991} eds. H Sato and T Nakamura (World Scientific, Singapore) 

1013-1032 (1992).

G W Gibbons and  H-J Pohle {\sl Nucl. Phys.} {\bf B 410} (1993) 117-142

G W Gibbons {\sl Classical and Quantum Gravity} {\bf 10} (1993) S575-S578 .

G W Gibbons {\sl  Int. J. Mod. Phys. } {\bf D 3} 61-70 (1994)

\medskip \item {[6]} J S Milson {\sl Ann Math} {\bf 104} (1976) 235-247 

\medskip \item {[7]} G Besson, G Courtois and S Gallot {\sl C R Acad Sci Paris} {\bf 319} (1994) 81-84 

{\sl Geometry and Functional Analysis} {\bf 5} (1995) 1016-1443

\medskip \item {[8]} W Boucher, in {\sl  Classical General Relativity}. edited by W B  Bonner, J N Islam and M A H MacCallum ( Cambridge University Press, Cambridge, England) (1983) 

H Friedrichs {\sl J Geom Phys} {\bf 3} 101 (1986)

\medskip \item {[9]} S Ding, Y Maeda and M Siino {\sl Phy Lett} {\bf B 345} (1995)
46-51 

S Ding, Y Maeda and M Siino preprint TIT/HEP-282/COSMO-52, gr-qc/9503026

\medskip \item {[10]} L F Abbot and S Deser {\sl Nucl Phys } {\bf B159} (1982) 76-96 

P Breitenlohner and D Z Freedman {\sl Ann Phys } { \bf 144} (1982) 249-276 

G W Gibbons, C M Hull and N P Warner {\sl Nucl Phys } { B 218} 173-190 (1983)
\medskip 

\medskip \item {[11]} M W Davis {\sl Proc Amer Math Soc } {\bf 93} (1985) 
325-328

\medskip \item {[12]} J G Ratcliffe and S T Tschantz {\sl Volumes of 
Hyperbolic Manifolds}  (Nashville preprint)

R Kellerhals {\sl Mathematical Intelligencer} {\bf 17} (1995) 21-30  

\medskip \item {[13]} G W Gibbons, S W Hawking and M J Perry {\sl Nucl Phys} {\bf B 138 } (1978) 141-150

\medskip \item {[14]} G. Robert Monpelier Thesis

\medskip \item {[15]} R L Bishop {\sl Notices Amer. Math Soc.} {\bf 10} 364 (1963)

\medskip \item {[16]} S W Hawking {\sl Phys Lett }{\bf 134} (1984) 403-404 

\medskip \item {[17]} M J Duff "The cosmological constant is possibly zero but the proof is probably wrong" CTP-TAMU 16-89 pp 403-408  of Strings 89 ed R Arnowitt, published by by World Scientific (1990)

\medskip \item {[18]} G W Gibbons {\sl Phys Lett} {\bf A 61 } (1977) 3-5

\medskip \item {[19]} S W Hawking {\sl Physica Scripta} {\bf T 15} (1987) 151

D N Page {\sl Phys Rev} { \bf D 34} (1986) 2267-2271

\medskip \item {[20]} A O Barvinsky, V P Frolov and A I Zelnikov {\sl Phys Rev} {\bf D 51} (1995) 1741-1763 

\medskip \item {[21]} R Penrose in {\sl The Nature of Space and Time} Princeton University Press (1996) 

\medskip \item {[22]} V D Dzhunushaliev {\sl prerint} gr-qc/9512014

\medskip \item {[23]} D Bleeker {\sl J Diff Geom }  {\bf 14} (1979) 599-608 

\medskip \item {[24]} S Tannno {\sl J Diff Geom} {\bf 14} (1979)  237-240

\medskip \item {[25]} S A Hayward, T Shiromizu and K Maeda 
{\sl Phys Rev} {\bf D49 } (1993) 2080-2085

\medskip \item {[26]} G Hayward and J Twamley {\sl Phys Lett} {\bf A 149} (1990) 84-90

\medskip \item {[27]} S V Matveev and A T Fomenko {\sl Russian Math. Surveys}
{\bf 43} 3-24(1988) 

\medskip \item {[28]} G W Gibbons {\sl Nucl. Phys.}{\bf B292} (1987) 784-792 

{\sl Nucl. Phys.} {\bf B310} (1988) 636 -642 

\medskip \item {[29]} R Calderwall, A Chamblin and G W Gibbons 
(in prepartion) 

\medskip \item {[30]} M Cvetic and S Griffies {\sl Phys Lett} {\bf B285} (1992)

G W Gibbons {\sl Nucl Phys} {\bf B 394} (1993) 3-20 

\medskip \item {[31]} M Lachieze-Rey and J.P. Luminet {\sl Phys Rep} {\bf 254} (1995) 135

\medskip \item {[32]} N J Cornish, D N Spergel and G D Starkman {\sl preprint} astro-ph/9601034

\medskip \item {[33]} V P Frolov and I D Novikov {\sl Phys Rev} {\bf D 42} (1tem {[34]} J D Barrow and D H Sonoda {\sl Phys Rep} {\bf 159} (1986) 1-49

\bye